\documentclass[aps,prl,reprint,superscriptaddress,showpacs]{revtex4-1}
\usepackage{color}
\usepackage{graphicx}
\usepackage[american]{babel}
\usepackage[utf8]{inputenc}
\usepackage[T1]{fontenc}
\usepackage{bbm,ae,aecompl}
\usepackage{amsmath}
\usepackage{amssymb}
\usepackage{amstext}
\usepackage{amsthm}
\usepackage{latexsym}
\usepackage{verbatim}
\usepackage{natbib}

\usepackage{ulem}   

\begin{document}

\title{Large temperature dependence of the number of carriers in Co-doped BaFe$_2$As$_2$}

\date{\today}
\pacs{79.60.-i, 71.18.-y, 71.30.-h}

\author{V. Brouet}

\author{Ping-Hui Lin}

\author{Y. Texier}

\author{J. Bobroff}

\affiliation{Laboratoire de Physique des Solides, Universit\'{e} Paris-Sud, UMR 8502, B\^at. 510, 91405 Orsay, France}

\author{A. Taleb-Ibrahimi}

\author{P. Le F\`evre}

\author{F. Bertran}
\affiliation{Synchrotron SOLEIL, L'Orme des Merisiers, Saint-Aubin-BP 48, 91192 Gif sur Yvette, France}

\author{M. Casula}
\affiliation{Institut de Min\'{e}ralogie et de Physique des Milieux condens\'{e}s,
Universit\'{e} Pierre et Marie Curie, CNRS, 4 place Jussieu, 75252 Paris, France}

\author{P. Werner}
\affiliation{Department of Physics, University of Fribourg, 1700 Fribourg, Switzerland}

\author{S. Biermann}
\affiliation{Centre de Physique Th\'{e}orique, Ecole Polytechnique, UMR 7644, 91128 Palaiseau Cedex, France}

\author{F. Rullier-Albenque}

\author{A. Forget}

\author{D. Colson}
\affiliation{Service de Physique de l'Etat Condens\'{e}, Orme des Merisiers, CEA Saclay, CNRS-URA 2464, 91191 Gif sur Yvette Cedex, France}

\begin{abstract}
Using angle-resolved photoemission spectroscopy, we study the evolution of the number of carriers in Ba(Fe$_{1-x}$Co$_x$)$_2$As$_2$ as a function of Co content and temperature. We show that there is a k-dependent energy shift compared to density functional calculations, which is large at low Co contents and low temperatures and reduces the volume of hole and electron pockets by a factor 2. This k-shift becomes negligible at high Co content and could be due to interband charge or spin fluctuations. We further reveal that the bands shift with temperature, changing significantly the number of carriers they contain (up to 50$\%$). We explain this evolution by thermal excitations of carriers among the narrow bands, possibly combined with a temperature evolution of the k-dependent fluctuations. 
\end{abstract}
 
\maketitle

Since the discovery of iron based superconductors, angle resolved photoelectron spectroscopy (ARPES) has given valuable information about their electronic structure \cite{KordyukReview,RichardReview}. In most cases, the measured spectra are in broad agreement with Density Functional Theory (DFT) calculations, after renormalization by a factor 2 to 3. This range of values is in good agreement with mass enhancements predicted by Dynamical Mean Field Theory (DMFT) \cite{YinNatureMat11,WernerNatPhys12}. More unusual is a shrinking of the hole and electron pocket sizes compared to DFT calculations, observed both by ARPES and de Haas-Van Alphen experiments \footnote{A.I. Coldea et al., Phys. Rev. Lett. 101, 216402 (2008) in LaFePO. Up to now, dHvA osciallations could not be observed in Co-doped BaFe$_2$As$_2$}. In a Fermi liquid (FL) picture, this corresponds to a down shift of the hole bands compensated by an up shift of the electron ones, as sketched in Fig. 1(a). As it depends on k, we refer to this effect as k-shift in the following. Its origin 
is so far unclear, but it seems associated with the strength of correlations. In P-substituted BaFe$_2$As$_2$, dHvA measurements reported that the maximum k-shift corresponds to the largest effective masses \cite{ShishidoPRL10}. In Ru-substituted BaFe$_2$As$_2$, both the k-shift and the renormalization weaken at large doping content \cite{BrouetRu,XuDingPRB12}. Ortenzi \textit{et al.} showed that interband interactions mediated by low energy bosons, such as spin fluctuations, can lead to such a k-shift \cite{OrtenziPRL09}. If this is indeed the case in pnictides, it contains important information about the interactions in these compounds and deserves a detailed study. 

We show here with ARPES in Co-doped BaFe$_2$As$_2$ that the k-shift is suppressed for high Co content, when the hole pockets fill up. We reveal in addition a very strong temperature (T) dependence of the electronic structure. Up to 10$\%$ Co, the electron pockets are expanding with increasing T by an amount as large as 50$\%$. A similar T dependence was reported recently for Ru-doped BaFe$_2$As$_2$ \cite{DhakaTemp}. At higher Co content (30$\%$), the electron pockets on the contrary shrink at high T. We show that this can be largely explained by thermal excitations in these narrow multiband systems. We present both simple models and calculations using the density of states (DOS) obtained by DFT within the Local Density Approximation (LDA) and by DMFT in order to simulate the effect of temperature within a FL framework. These simulations allow to account for about half the experimental shift. Part of the expansion of the electron pockets could also be due to the weakening of the k-shift, as proposed in \cite{BenfattoPRB11} for the model of interband interactions, or to an evolution of correlation effects. 

Single crystals of Ba(Fe$_{1-x}$Co$_x$)$_2$As$_2$ were grown using a FeAs self-flux method and studied in detail by transport measurements \cite{RullierAlbenquePRL09}. ARPES experiments were carried out at the CASSIOPEE beamline at the SOLEIL synchrotron, with a Scienta R4000 analyzer, an angular resolution of $0.2^{\circ}$ and an energy resolution of $15$ meV. All measurements were perfomed at 34 eV, except otherwise mentioned. DFT band structure calculations were performed using the Wien2k package \cite{Wien2k} at the experimental structure of BaFe$_2$As$_2$. To model Co doping, we used the virtual crystal approximation (VCA), where Co induces a global rigid shift of the entire band structure. For the LDA+DMFT calculation, we used the technology of ref \cite{WernerNatPhys12}, which fully takes into account dynamical screening effects from first principles. 

\begin{figure*}[tbp]
\centering
\includegraphics[width=0.9\textwidth]{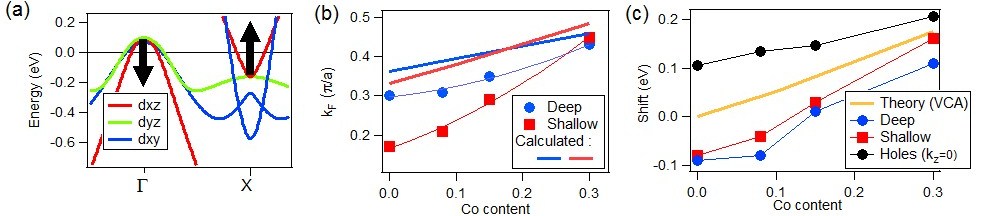}
\caption{(a) Calculated LDA band structure along $\Gamma$X at $k_z$=0 and x=0. Colors indicate the main orbital character. Arrows indicate the direction of the k-shift observed at low T in experiments. (b) $k_F$ measured as a function of Co content for the two electron bands, deep and shallow. The thin lines are guides for the eye. Solid lines are  $k_F$ obtained in VCA DFT calculations for $k_z$ corresponding to experimental conditions \cite{sup}. 
(c) Shifts compared to LDA calculations at x=0 needed to obtain the experimental $k_F$. Solid line indicates the expectations from the VCA calculations.}
\label{NewFig1}
\end{figure*}

Recently, we have shown that the electron pocket around X is an ellipse formed by a \lq\lq{}deep\rq\rq{} electron band of $d_{xy}$ symmetry along the long axis and a \lq\lq{}shallow\rq\rq{} electron band of $d_{xz}$/$d_{yz}$ symmetry along the short axis \cite{BrouetPRB12}. In Fig. 1(b), we report $k_F$ for these two bands as a function of Co doping at 20 K (150 K for BaFe$_2$As$_2$ to avoid the magnetic state). In the supplementary information \cite{sup}, we show the full dispersions used to extract these values. The pocket expands with Co content, as expected from the effective electron doping of the bands \cite {BrouetPRB09,ThirupathaiahPRB10,LiuPRB11_Co,NeupaneDingPRB11}. For small Co contents, the values are significantly smaller in the experiment than in the DFT VCA calculations, shown as solid lines. This reflects the up shift of the electron bands sketched in Fig. 1(a) and described in the introduction. Interestingly, the shift also increases significantly the anisotropy of the pocket near x=0 doping.

In Fig. 1(c), we report as a function of Co content the shifts with respect to the LDA bands calculated at x=0 needed to obtain the measured $k_F$. As $k_F$ is uniquely determined by the number of carriers in one pocket, this shift value is independent of the band renormalization and well defined. We found similar shift values for different $k_z$ \cite{BrouetPRB12}. Compared to the rigid shift found in the DFT VCA calculation (orange solid line), the bare shift is nearly 100 meV at low dopings, but it vanishes at higher dopings. In Table I, we quote the shift at x=0, as well as the renormalization extracted from the dispersion \cite{sup}. 


It is more difficult to perform a similar analysis for the hole pockets, because the shifts and renormalization values are strongly $k_z$-dependent (see Table I and \cite{sup}). Such a deformation of the band structure is not surprising considering that the bands must shift in opposite directions at $\Gamma$ and X. Fig. 1(a) recalls for example that the $d_{yz}$ hole band must be degenerate at X with the $d_{xz}$ electron band. As this point is fixed in energy along $k_z$ by symmetry, while the hole bands shift quite strongly, this implies some modifications of the hole pockets $k_z$ dependence. We report in Fig. 1(c) the shift of the calculated $d_{xz}$/$d_{yz}$ bands needed to describe the top of the hole band at $k_z$=0. Like for electron pockets, the k-shift disappears quite precisely when the hole pockets become completely filled, near x=0.2 \cite {BrouetPRB09}, suppressing the possibility of adjusting the number of carriers by relative shifts of the bands.

\begin{table}[b]
\caption{\label{''Table 1''} Shifts and effective masses with respect to LDA bands calculated at x=0 \cite{sup}.}
\begin{tabular}{|c|c|c|c|c|}  
\hline
      & Holes $k_z$=0 & Holes $k_z$=1  & Deep elec.  & Shallow elec.  \\
    &                          &  (outer)            &  ($d_{xy}$)       & ($d_{xz}$/$d_{yz}$) \\
  \hline
 shift (meV) &110&30&-90&-80\\
m*/m$_b$ & 1.5 & 3 & 2 & 2.2\\
   \hline
\end{tabular}
\end{table}

The temperature dependence should give interesting and complementary information to explore the nature of the k-shift. If related to spin fluctuations, one could expect interband interactions to weaken at high T, as the strength of spin fluctuations decreases \cite{BenfattoPRB11}. In Fig. 2, we show T dependent measurements from 20 to 300 K for 8$\%$ and 30$\%$ Co contents. The full dispersion images are shown in the supplementary information \cite{sup}. In each case, we took care to check that the effect was reversible, as there is often some evolution with temperature cycles (broadening and/or loss of intensity). In Fig. 2(a) and 2(d), we show the spectral intensity of the shallow band integrated in a 10 meV window around the chemical potential $\mu$. The distance at a given temperature between the two maxima defines 2$k_F$. White dashed lines indicate its values at low T to emphasize the evolution at higher T. As reported by red squares in Fig. 2(b) and 2(e), $k_F$ seems roughly constant up to 150K and then increases at 8$\%$ Co and decreases at 30$\%$ Co. For the other bands, there is also an evolution of $k_F$ with T, which is reported in the same figures. For Co 30$\%$, the hole pockets are filled and small electron pockets appear at $\Gamma$ \cite{sup,LiuPRB11_Co}. The opposite evolution of $k_F$ for the shallow electron pocket for the two dopings evidences that the evolution depends on the nature of the carriers at $\mu$. In fact, these changes of  $k_F$ correspond to a shift of the whole bands \cite{sup}. In Fig. 2(c) and 2(f), we report these shifts as a function of T. For Co 8$\%$, they are remarkably similar, indicating that the evolution in temperature is mainly driven by a shift of the $\mu$ within the electronic structure.    

\begin{figure}[tbp]
\centering
\includegraphics[width=0.5\textwidth]{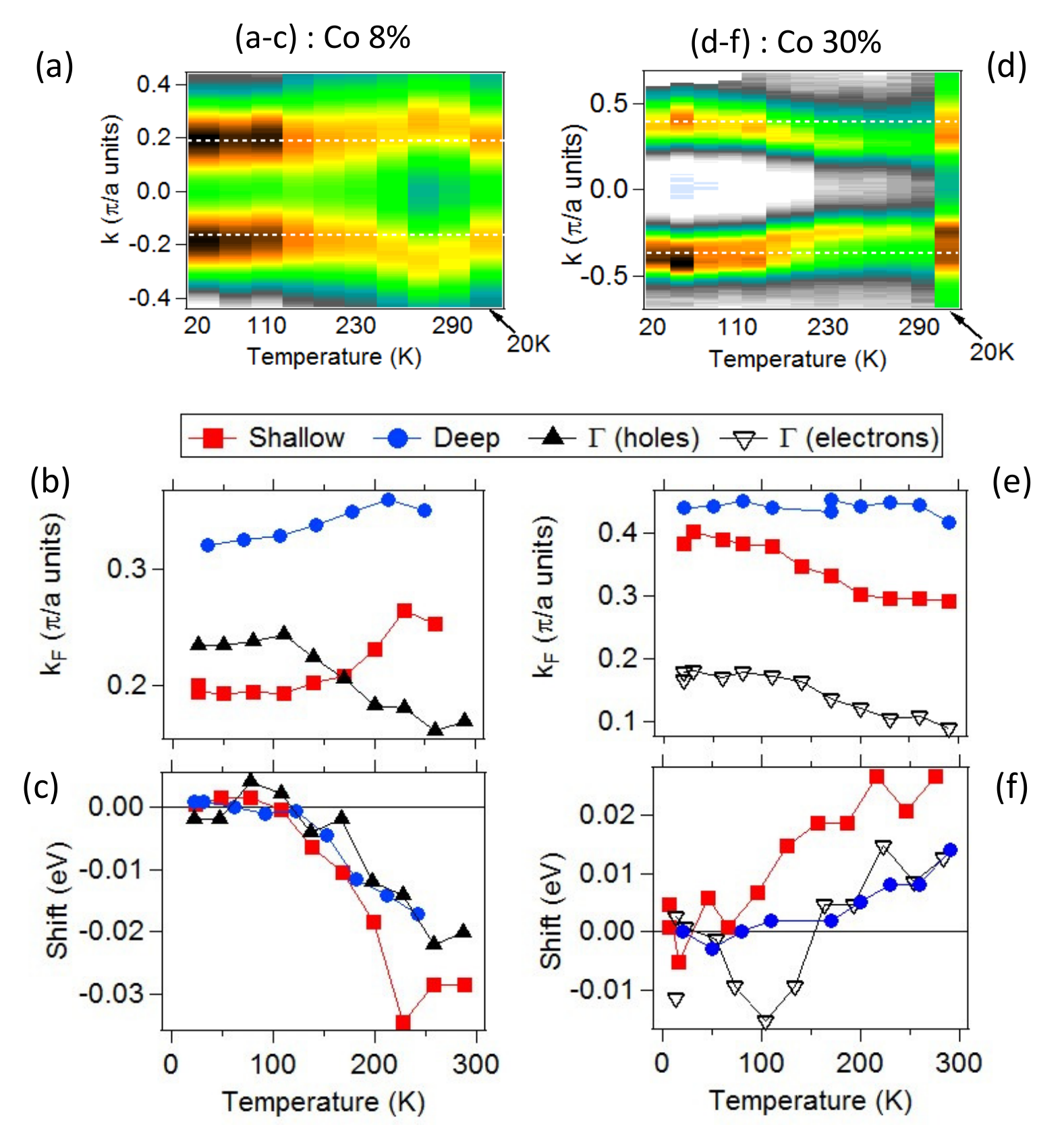}
\caption{Temperature dependence for 8$\%$ Co (a-c) and 30$\%$ Co (d-f). (a,d) Spectral weight integrated over 10meV around $\mu$ for the shallow electron band. White dashed line indicate $k_F$ at T=0. The last row (20 K) was measured after cooling back from 300K. (b,e) $k_F$ as a function of T for the two electron bands at X (shallow and deep) and for one band at $\Gamma$ (holes for x=0.08, electrons for x=0.3). (c,f) Shift as a function of T for the different bands. }
\label{Fig3}
\end{figure}


In Fig. 3(a), we give a simple sketch of the Co 8$\%$ band structure with realistic parameters : a hole band whose top is 10 meV above $E_F$ (in experiment, it is even below $E_F$ for a large range of k$_z$\cite{sup}) and an electron band whose bottom is 50 meV below $E_F$, as for the shallow electron band. We assume a constant DOS $\rho(E)$ and choose it so that both bands contain the same number of carriers $n_0$ at T=0. At 300K, the Fermi edge (thin line) extends over the top or bottom of the bands. If the chemical potential $\mu$ did not move, the number of carriers $n=\int \!  \rho(E)f(E) \, \mathrm{d}$E would be quite different from T=0, namely 1.07$n_0$ for electrons and 2.33$n_0$ for holes in this example. To keep the number of carriers constant, $\mu$ has to shift to a higher position that will reduce the number of excited carriers. As the imbalance for holes is much larger than for electrons, $\mu$ stabilizes a different value of $n$ compared to T=0. A simple calculation shows that the new number of carriers would be 1.38$n_0$ [Fig. 3(b)] and $\mu$ should shift by 17 meV at 300 K to ensure $n_e$=$n_h$ [blue line in Fig. 3(d)]. Note that, despite the fact that the hole bands are now completely below $E_F$, as in our measurement, they do contain carriers. Similarly, the fact that $k_F$ at Co 8$\%$ decreases for holes and increases for electrons does not violate the conservation of the number of carriers. Although very rough, this model gives the right order of magnitude for the observed shift, supporting the idea that this effect plays a leading role. 
\begin{figure}[tbp]
\centering
\includegraphics[width=0.5\textwidth]{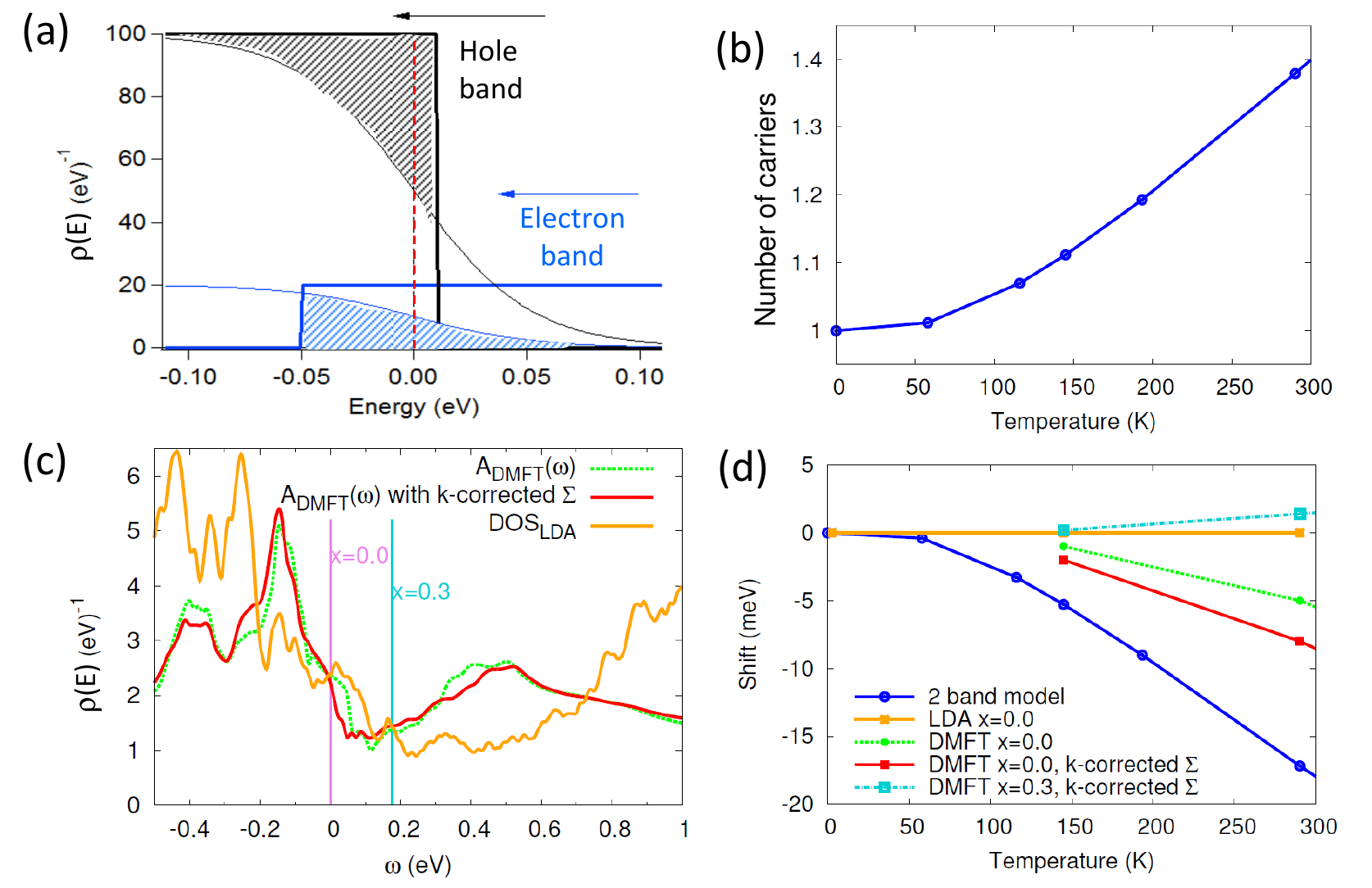}
\caption{(a) Thick lines : DOS considered for the 2 bands model. The Fermi edge is sketched at 300K, dashed areas indicate the occupied parts for holes and electrons before shifting the chemical potential $\mu$. Arrows indicate the directions in which the bands will shift with increasing T with respect to $\mu$. (b) Number of carriers (holes or electrons) obtained as a function of temperature for the 2 bands model. (c) DOS calculated in LDA, LDA+DMFT and LDA+DMFT including a k-shift. The Fermi levels for x=0 and x=0.3 are indicated by vertical lines. (d) Shift of $\mu$ obtained within band  theory from the DOS of various models (see text and caption). }
\label{Fig4}
\end{figure}
To go further, we now consider in Fig. 3(c) a realistic DOS for BaFe$_2$As$_2$. As in our simple model, it is the occurrence of fine structures within $4k_BT$ of the Fermi level that will induce a shift of $\mu$. Indeed, within FL theory, one expects $\delta\mu(E) \propto T^2 \rho^\prime(E)/\rho(E)$ \footnote{N.W Ashcroft, N.D. Mermin, Solid State Physics, Brooks Cole (1976)}. In the LDA DOS, the peaks are too far from $E_F$ to induce any sizable shift and $\mu$ remains constant in Fig. 3(d). As a next step, we use the DMFT spectral function calculated at 150 K as an approximation for the DOS. The renormalization of the bands bring the peaks closer to $E_F$ and a sizable shift is detectable at 300K in Fig. 3(d). This shift even increases when the bands extrema are moved closer to $E_F$ by introducing arbitrarily a k-dependent shift simulating that observed experimentally \footnote{The DMFT spectral function has been evaluated by means of the self-energy analytic continuation and an ultra-dense k-integration (20x20x20 k-point grid). Whenever specified, the self-energy has been corrected by the simple functional form $\epsilon (1 - 2 \cos(k_x) \cos(k_y))$, with $\epsilon$ set to 0.08 eV to reproduce the experimental positions of the bands at $\Gamma$ and X. This static k-shift correction has been applied rigidly to all orbitals. Doping has been introduced by VCA, as in the LDA calculations.}. For x=0.3, $E_F$ moves to a region of opposite slopes inducing a change of sign in the shift, as observed in Fig. 2(f). Even if the values are somewhat smaller than in the experiment, this captures the qualitative behavior very well. As will be discussed in the conclusions, a T dependence of the k-shift could also contribute to these effects.

In Fig. 4, we finally compute the number of electrons as a function of temperature and doping. Thanks to the knowledge of the 3D shape of the electron pockets \cite{BrouetPRB12}, this can be reliably obtained by combining the $k_F$ measured for the deep and shallow bands and integrating over $k_z$ \footnote{Strictly speaking, this computation based on $k_F$ is only valid at T=0. However, the electron pockets are sufficiently deep to make the T correction negligible.}. In the right part of the figure, we report the change of the number of electrons as a function of T. For 8$\%$ Co, this leads to an increase by 50$\%$ between low and high temperature. This order of magnitude is in good agreement with the simple model of Fig. 3. For BaFe$_2$As$_2$ above the magnetic transition, the effect is also present and even slightly larger. This rules out that the anomalous temperature dependence could be due to perturbations associated with the presence of Co. For 15$\%$ Co, the effect is somewhat reduced. For 30$\%$ Co, the number of electrons at X decreases with increasing T [see $k_F$ in Fig. 2(e)], but it is difficult to evaluate the number of electrons at $\Gamma$, so that we only indicate the low T value \footnote{The small electron pocket at $\Gamma$ cannot be detected at high T \cite{sup}, but certainly contains carriers. Moreover, Fig. 2(f) suggests some complex orbital dependent T evolution.}. The left part of Fig. 4 displays the number of electrons as a function of Co content. This emphasizes the amplitude of the change in temperature compared to that with doping. It also shows that the T dependence tends to reduce the deficit of electrons present at low T compared to the calculation. For 30$\%$ Co, the hole pockets are completely filled and the number of electrons is almost that expected by stoichiometry. 

Interestingly, this T dependence of the number of electrons is strikingly similar to the one deduced from Hall measurements \cite{RullierAlbenquePRL09}, although a bit smaller quantitatively. This strongly supports the interpretation of both ARPES and transport based on a change of the number of electrons with T. As a result, much care should be taken to extract the genuine T dependence of the scattering times from resistivity measurements. Another quantity that exhibits a quite similar and anomalous T dependence is the spin susceptibility $\chi$, as measured by NMR for example \cite{NingJPHysSocJapan09}. In this case however, $\chi$ increases with temperature but decreases with electron doping. Of course, $\chi$ should not solely depend on the number of electrons, but also on the number of holes, for example. In any case, it will be important to consider this T dependent number of carriers to reinvestigate these properties \cite{KorshunovPRL09}. 

\begin{figure}[tbp]
\centering
\includegraphics[width=0.45\textwidth]{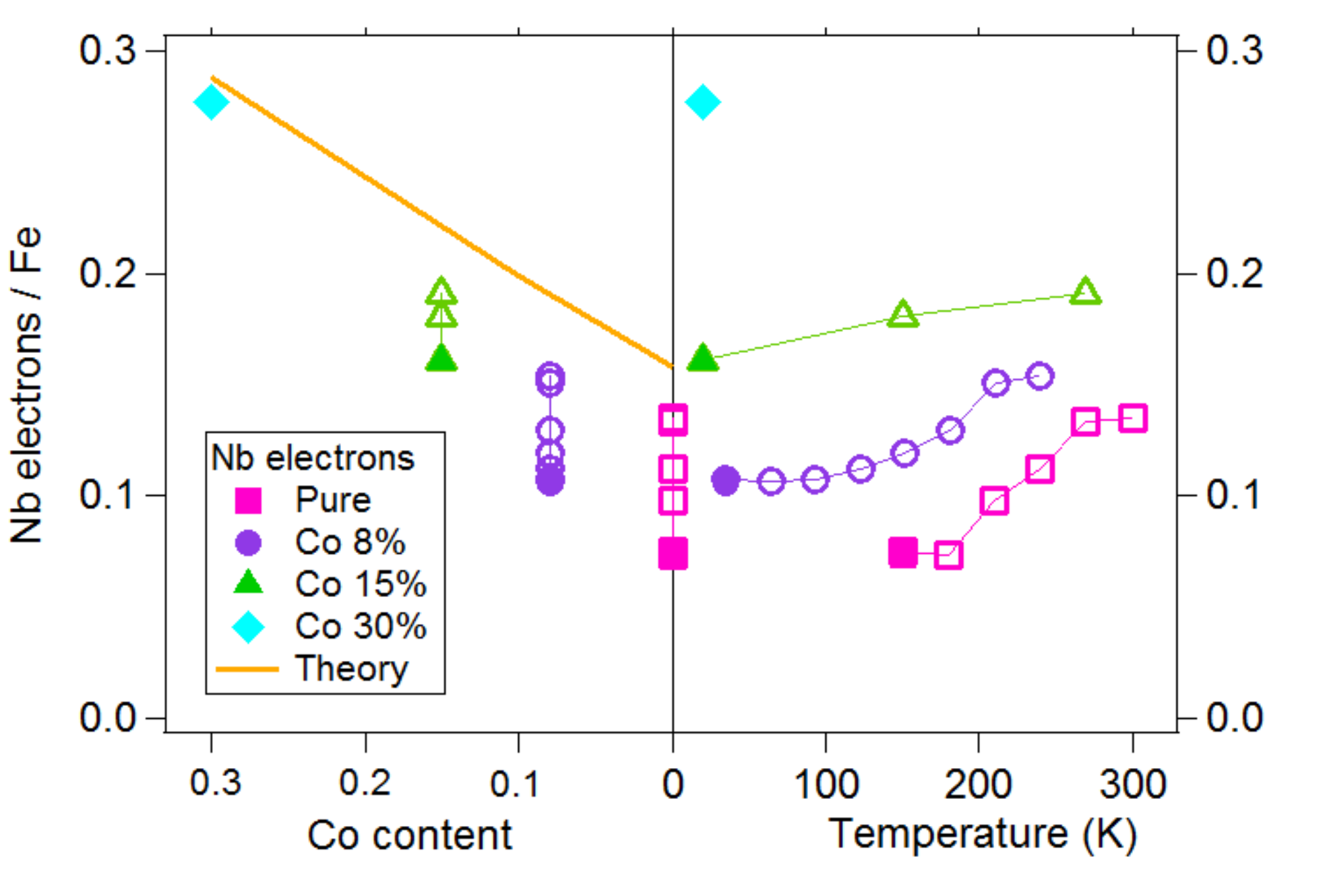}
\caption{Number of electrons as a function of Co content for different temperatures (left part) and as a function of T for different Co content (right part). The lowest T is shown as filled symbol. The orange thick line indicates the number of electrons expected from LDA calculation.}
\label{Fig4}
\end{figure}

Finally, our study reveals peculiar and so far largely unexplored consequences of the multiband nature of iron-based superconductors. First, there is a shift compared to band calculations that is k-dependent and leads to the shrinking of the Fermi Surface for low Co content. This k-shift has a dramatic effect as it reduces the total number of carriers by a factor of 2 for x=0. It also automatically leads to more elliptic electron pockets and more 3D hole pockets than in the LDA band calculation, as we observe by ARPES. These k-shifts weaken with doping, as the hole bands fill up, which is consistent with the picture based on interband couplings \cite{OrtenziPRL09,BenfattoPRB11}. It would be very interesting to obtain a full theoretical description of this effect. It has not been seen in DMFT based electronic structure calculations up to now, possibly because it requires an extremely high energy precision to be resolved (of the order of 10 meV). Alternatively, one may speculate that it is due to non-local self-energy effects not included at the DMFT level. A very recent GW study may be interpreted in this sense \cite{Tomczak2012}. Independently of their origin, an important consequence of these k-shifts is that they make the whole band structure much more sensitive to temperature than what would be expected in most calculations. Indeed the most striking finding of our work is a strong dependence of the electronic structure on temperature, confirming and generalizing the finding of a previous ARPES study \cite{DhakaTemp}. The resulting large T dependence of the number of carriers must impact all other electronic properties. There is already good evidence that transport probes a similar evolution \cite{RullierAlbenquePRL09}. We show that thermal excitations within the FL picture partly explain the effect. A full theoretical description will probably requires a complete understanding of the origin of the k-shift, so that its T dependence can be controlled.

\section{Acknowledgements}
We thank Jan Tomczak for useful discussions. Financial support from the ANR \lq\lq{}Pnictides\rq\rq{} is acknowledged.

\bibliography{Bib}

\end{document}